\documentclass[]{JHEP3}
\usepackage{epsfig,multicol,bbm}

\newcommand\fverb{\setbox\fverbbox=\hbox\bgroup\verb}
\newcommand\fverbdo{\egroup\medskip\noindent%
                        \fbox{\unhbox\fverbbox}\ }
\newcommand\fverbit{\egroup\item[\fbox{\unhbox\fverbbox}]}
\newbox\fverbbox

\title{Extra dimensions in 
${\gamma\gamma \to \gamma\gamma}$ process  at the CERN-LHC}

\author{S. Ata\u{g}\\
Department of Physics, Faculty of Sciences,
Ankara University, 06100 Tandogan, Ankara, Turkey\\
E-mail:\email{atag@science.ankara.edu.tr}}

\author{S.C. \.{I}nan \\
Department of Physics, Cumhuriyet University,
58140, Sivas, Turkey\\
E-mail:\email{sceminan@cumhuriyet.edu.tr}}

\author{\.{I}. \c{S}ahin \\
Department of Physics, Zonguldak Karaelmas University, 
67100 Zonguldak, Turkey \\
E-mail:\email{inancsahin@karaelmas.edu.tr}} 

\abstract{Potential of the LHC  to explore the phenomenology of the 
Kaluza-Klein (KK) tower of gravitons in the scenarios of the   
Arkani-Hamed, Dimopoulos and Dvali(ADD) model
and Randall-Sundrum (RS) model is discussed via the process 
${\gamma\gamma \to \gamma\gamma}$ including the Standard Model 
one loop diagram. The improved constraints on model parameters 
have been obtained  compared to the LEP and Tevatron sensitivity.}

\keywords{Phenomenology of Large extra dimension}

\begin{document}

\section{Introduction}
Central detectors of the Large Hadron Colliders (LHC) experiments 
ATLAS and CMS at CERN have a pseudorapidity $\eta$ coverage 
2.5 for tracking system and 5.0 for calorimetry. In spite of that,
a certain amount of particles and momentum flow take place in 
the very forward directions and are not detected by these detectors.
Furthermore, there are  contributions to the measured cross sections 
from elastic scattering and ultraperipheral collisions. 
For better understanding of the
physics from very forward region  one needs  additional equipments.
ATLAS and CMS collaborations developed a program of forward physics
with extra detectors located in a region nearly 100m-400m from the
interaction point \cite{royon}.
These are called  forward detectors which will be installed
very close  to the beamline. 
In connection with these new equipments 
a research plan has been organized  to investigate   
soft and hard diffraction,
high energy photon induced interactions,
low-x dynamics with forward jet studies,
large rapidity gaps between forward jets,
and luminosity monitoring \cite{royon,khoze, schul}.
One of the main features of these forward
detectors is to tag the protons  with some energy fraction loss
$\xi =E_{loss}/E_{beam}$.
Based on the ATLAS and CMS working conditions
forward detectors will be located in  positions 
from interaction point to have an overall acceptance
$0.0015<\xi<0.5$ \cite{royon2,albrow}.
Larger $\xi$ is created  if the  forward detectors are installed
closer to the interaction points.
Using the forward detectors it is possible to produce 
high energy photon induced interaction with exclusive
two particle  final states such as photons or leptons.  
In this work we are interested in the two photons 
in the  final states.
Almost real photons are emitted by each proton 
and  interact each other to provide two photons
$\gamma\gamma \to \gamma\gamma$.
Forward detectors will detect the energy loss of the 
deflected protons  but 
the  central detector will identify  the final photons
with rapidity $|\eta|<2.5$ and $p_{T}>(10-20)$ GeV. 
Emitted  photons by the protons with small angles
have  a spectrum depending on virtuality $Q^{2}$ and energy 
$E_{\gamma}$. This is described by the equivalent photon 
approximation \cite{budnev,baur}.  The spectrum differs 
from the pointlike electron positron case 
by taking into account the electromagnetic form 
factors 

\begin{eqnarray}
dN=\frac{\alpha}{\pi}\frac{dE_{\gamma}}{E_{\gamma}}
\frac{dQ^{2}}{Q^{2}}[(1-\frac{E_{\gamma}}{E})
(1-\frac{Q^{2}_{min}}{Q^{2}})F_{E}+\frac{E^{2}_{\gamma}}{2E^{2}}F_{M}]
\end{eqnarray}

where

\begin{eqnarray}
Q^{2}_{min}=\frac{m^{2}_{p}E^{2}_{\gamma}}{E(E-E_{\gamma})}, 
\;\;\;\; F_{E}=\frac{4m^{2}_{p}G^{2}_{E}+Q^{2}G^{2}_{M}}
{4m^{2}_{p}+Q^{2}} \\
G^{2}_{E}=\frac{G^{2}_{M}}{\mu^{2}_{p}}=(1+\frac{Q^{2}}{Q^{2}_{0}})^{-4}, 
\;\;\; F_{M}=G^{2}_{M}, \;\;\; Q^{2}_{0}=0.71 \mbox{GeV}^{2}
\end{eqnarray}  

Here E is the energy of the proton beam which is related to the 
photon energy by $E_{\gamma}=\xi E$ 
and $m_{p}$ is the mass of the proton.
The magnetic moment of the proton is $\mu^{2}_{p}=7.78$, $F_{E}$ and 
$F_{M}$ are functions of the electric and magnetic form factors.
The integration of the cross section 
$d\sigma_{\gamma\gamma \to 
\gamma\gamma}$  should be made over the photon spectrum  

\begin{eqnarray}
d\sigma=\int{\frac{dL^{\gamma\gamma}}{dW}} 
d\sigma_{\gamma\gamma \to \gamma\gamma}(W)dW
\end{eqnarray} 

where the effective photon luminosity $dL^{\gamma\gamma}/dW$ 
is given by  

\begin{eqnarray}
\frac{dL^{\gamma\gamma}}{dW}=\int_{Q^{2}_{1,min}}^{Q^{2}_{max}}
{dQ^{2}_{1}}\int_{Q^{2}_{2,min}}^{Q^{2}_{max}}{dQ^{2}_{2}}
\int_{y_{min}}^{y_{max}}
{dy \frac{W}{2y} f_{1}(\frac{W^{2}}{4y}, Q^{2}_{1}) 
f_{2}(y,Q^{2}_{2})}.
\end{eqnarray}
with 

\begin{eqnarray}
y_{min}=\mbox{MAX}(W^{2}/(4\xi_{max}E), \xi_{min}E), \;\;\;
y_{max}=\xi_{max}E, \;\;\;
f=\frac{dN}{dE_{\gamma}dQ^{2}}.
\end{eqnarray}
Here W is the invariant mass of the two photon system 
$W=2E\sqrt{\xi_{1}\xi_{2}}$ and maximum virtuality is   
$Q^{2}_{max}=2$ $\mbox{GeV}^{2}$.
In Fig.\ref{fig1} effective $\gamma\gamma$ luminosity 
is shown as a function of the invariant mass of the two 
photon system where pp energy $\sqrt{s}=14$ TeV is taken.

\smallskip
\FIGURE{\epsfig{file=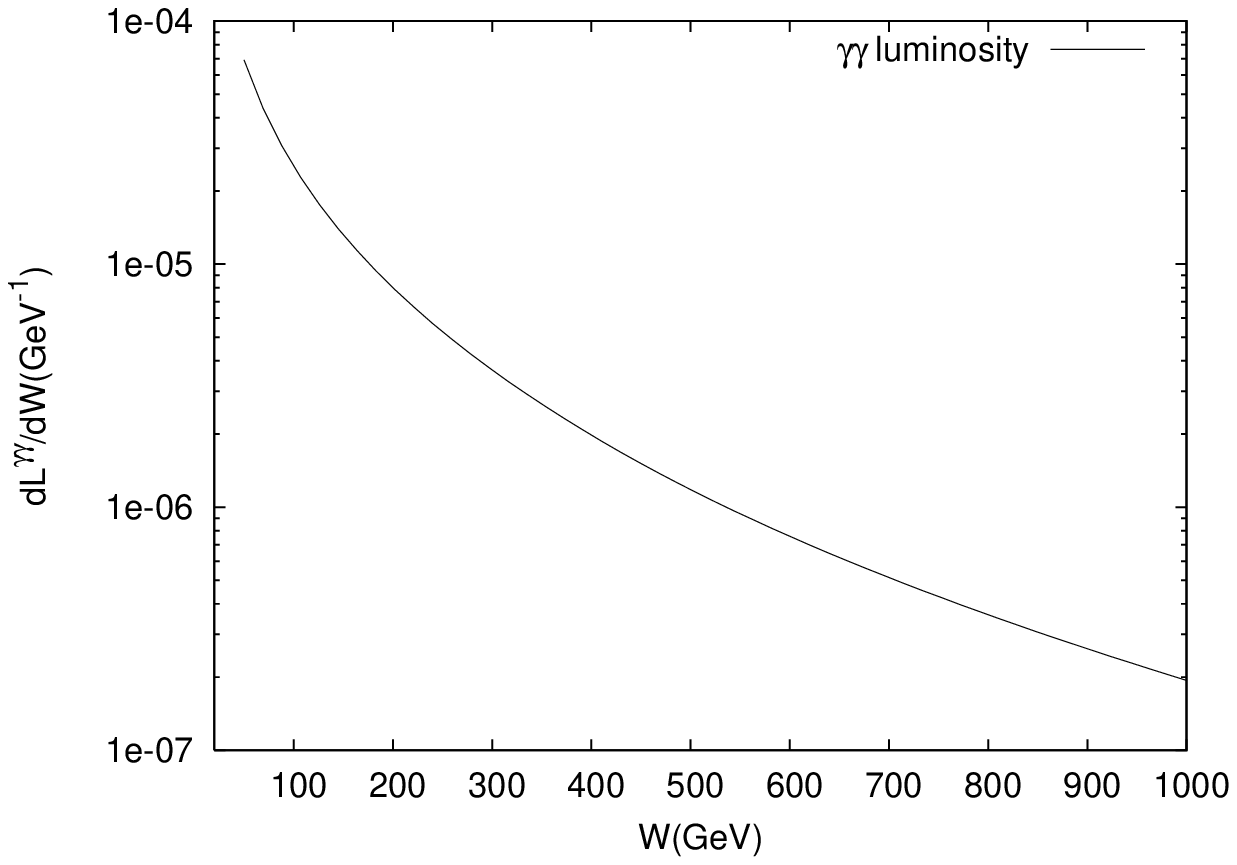}
\caption{Effective $\gamma\gamma$ luminosity as a function
of the invariant mass of the two photon system for
$0.0015<\xi<0.5$. pp  energy $\sqrt{s}=14$ TeV is taken.}
\label{fig1}}

It is interesting to search for new physics 
with invariant two photon mass $W > 1$ TeV
by  the photon-induced two photon final states 
with available luminosity. 
In this work, we explore the phenomenology of extra dimensions 
in the framework of  the Arkani-Hamed, 
Dimopoulos and Dvali(ADD) model of 
large extra dimensions and Randall-Sundrum (RS) model 
of warped extra dimensions via the photon induced process 
$pp \to p\gamma\gamma p$ at the LHC. We also summarize the 
Standard Model  ${\gamma\gamma \to \gamma\gamma}$ loop process 
in its own right and as a background. $\sqrt{s}=14$ TeV is considered 
when the proton-proton center of mass energy is needed. 

\section{Standard Model ${\gamma\gamma \to \gamma\gamma}$ 
one-loop Process}

In this section, we review the properties of the 
helicity amplitudes of the  one-loop 
${\gamma\gamma \to \gamma\gamma}$ process and provide the 
SM cross sections at LHC using the steps taken in the previous 
section. An experiment involving 
$\gamma\gamma$ scattering at high energies has not been done yet. 
In this respect, the precise determination of this 
cross section have a special importance 
to test the renormalization procedure 
of the parts of the SM containing W gauge bosons.
Furthermore, this loop process becomes a background for 
new physics searches through $\gamma\gamma$ scattering.
        
One-loop diagrams involve charged fermions and W bosons. 
For the $\sqrt{s_{\gamma\gamma}}>200-250$ GeV W boson contribution 
is dominant over the lepton and quark contributions \cite{jikia}. 
This  nice  property allows to separate the  contributions of the 
fermions and W bosons in the cross section. There are 
sixteen helicity amplitudes contributing to the process.
Fermion contribution with mass $m_{f}$ and charge $Q_{f}$ 
 to the helicity amplitudes can be given simply 
by neglecting the terms like $m_{f}^{2}/\hat{s}$, $m_{f}^{2}/\hat{t}$
and $m_{f}^{2}/\hat{u}$ 

\begin{eqnarray}
\frac{1}{\alpha^{2}Q_{f}^{4}}M_{++++}^{f}(\hat{s},\hat{t},\hat{u})
=-8-8(\frac{\hat{u}-\hat{t}}{\hat{s}})Ln({\frac{\hat{u}}{\hat{t}}})
 -4(\frac{\hat{t}^{2}+\hat{u}^{2}}{\hat{s}^{2}})
[Ln({\frac{\hat{u}}{\hat{t}}})Ln({\frac{\hat{u}}{\hat{t}}})+\pi^{2}],
\end{eqnarray}    

\begin{eqnarray}
M_{+++-}^{f}(\hat{s},\hat{t},\hat{u})\simeq 
M_{++--}^{f}(\hat{s},\hat{t},\hat{u})\simeq 8\alpha^{2}Q_{f}^{4}.
\end{eqnarray}
where invariant Mandelstam  variables are defined as 
$\hat{s}=(p_{1}+p_{2})^{2}$, $\hat{t}=(p_{1}-p_{3})^{2}$ and 
$\hat{u}=(p_{2}-p_{3})^{2}$.
The remaining amplitudes can be related to those given by the 
equations above   by parity and Bose symmetry.

It is possible to get simple expressions for W contribution 
when the terms like  $m_{W}^{2}/\hat{s}$, $m_{W}^{2}/\hat{t}$
and $m_{W}^{2}/\hat{u}$ are neglected

\begin{eqnarray}
\frac{1}{\alpha^{2}}M_{++++}^{W}(\hat{s},\hat{t},\hat{u})&&=-16i\pi 
[\frac{\hat{s}}{\hat{t}}Ln(\frac{-\hat{t}}{m_{W}^{2}})
+\frac{\hat{s}}{\hat{u}}Ln(\frac{-\hat{u}}{m_{W}^{2}})]\nonumber \\ 
&&+12+12(\frac{\hat{u}-\hat{t}}{\hat{s}})Ln(\frac{\hat{u}}{\hat{t}})
\nonumber\\
&&+16(1-\frac{3\hat{t}\hat{u}}{4\hat{s}^{2}})
[Ln(\frac{\hat{u}}{\hat{t}})Ln(\frac{\hat{u}}{\hat{t}})+\pi^{2}]
\nonumber \\
&&+16[\frac{\hat{s}}{\hat{t}}Ln(\frac{\hat{s}}{m_{W}^{2}})
Ln(\frac{-\hat{t}}{m_{W}^{2}})+
\frac{\hat{s}}{\hat{u}}Ln(\frac{\hat{s}}{m_{W}^{2}})
Ln(\frac{-\hat{u}}{m_{W}^{2}})\nonumber \\
&&+\frac{\hat{s}^{2}}{\hat{t}\hat{u}}Ln(\frac{-\hat{t}}{m_{W}^{2}})
Ln(\frac{-\hat{u}}{m_{W}^{2}})],
\end{eqnarray}

\begin{eqnarray}
\frac{1}{\alpha^{2}}M_{+-+-}^{W}(\hat{s},\hat{t},\hat{u})&&=
-i\pi[12(\frac{\hat{s}-\hat{t}}{\hat{u}})
+32(1-\frac{3\hat{t}\hat{s}}{4\hat{u}^{2}})
Ln(\frac{\hat{s}}{-\hat{t}})\nonumber \\
&&+16\frac{\hat{u}}{\hat{s}}Ln(\frac{-\hat{u}}{m_{W}^{2}})  
+16\frac{\hat{u}^2}{\hat{t}\hat{s}}Ln(\frac{-\hat{t}}{m_{W}^{2}})]
+12 \nonumber \\
+&&12(\frac{\hat{s}-\hat{t}}{\hat{u}})Ln(\frac{\hat{s}}{-\hat{t}})
+16(1-\frac{3\hat{t}\hat{s}}{4\hat{u}^{2}})
Ln(\frac{\hat{s}}{-\hat{t}})Ln(\frac{\hat{s}}{-\hat{t}})
\nonumber \\
&&+16[\frac{\hat{u}}{\hat{t}}Ln(\frac{-\hat{u}}{m_{W}^{2}})
Ln(\frac{-\hat{t}}{m_{W}^{2}})
+\frac{\hat{u}}{\hat{s}}Ln(\frac{-\hat{u}}{m_{W}^{2}})
Ln(\frac{\hat{s}}{m_{W}^{2}})\nonumber \\
&&+\frac{\hat{u}^{2}}{\hat{t}\hat{s}}Ln(\frac{-\hat{t}}{m_{W}^{2}})
Ln(\frac{\hat{s}}{m_{W}^{2}})],
\end{eqnarray}

\begin{eqnarray}
M_{+-+-}^{W}(\hat{s},\hat{t},\hat{u})=
M_{-+-+}^{W}(\hat{s},\hat{t},\hat{u}),
\end{eqnarray}

\begin{eqnarray}
M_{+++-}^{W}(\hat{s},\hat{t},\hat{u})\simeq
M_{++--}^{W}(\hat{s},\hat{t},\hat{u})\simeq -12\alpha^{2}.
\end{eqnarray}

In the W contribution, one should note that the dominant terms 
arise from the imaginary part of the helicity amplitude for the 
energy region satisfying the approximation $m_{W}^{2}<<\hat{s}$.
For $m_{W}^{2}>>\hat{s}$ contribution of W boson decreases and 
fermion contribution drastically increases and much larger than the 
W contribution. The top quark contribution is neglected above
because it is much lower 
than the light quark  and W contribution in both regions . 
Fig.\ref{fig2} shows the contributions of fermions and W boson to the
cross sections $pp \to p\gamma\gamma p$ as a function of 
invariant mass of the two photon system for the energy region of 
200-500 GeV and for the rapidity $|\eta|<2.5$. Dominance of 
the W contribution is clear in this energy region.
The total cross section of fermion contributions  for 
W=20-80 GeV, $|\eta|<2.5$ is 0.723 fb 
and W boson contribution for W=200-500 GeV,
$\sqrt{|\hat{t}|}>200 $ GeV, $\sqrt{|\hat{u}|}>200$ GeV, 
$|\eta|<2.5$  is 0.074 fb. In both cases we take into account 
$0.0015<\xi<0.5$. 
Rapidity distributions of W-boson and fermion contributions 
to the cross section are provided in Fig.\ref{fig3} for two 
energy scales of two photon system 20-80 GeV and 200-500 GeV.  
As it is seen,
the curves which  belong to the fermions and W boson 
have different behaviour as rapidity changes.

\FIGURE{\epsfig{file=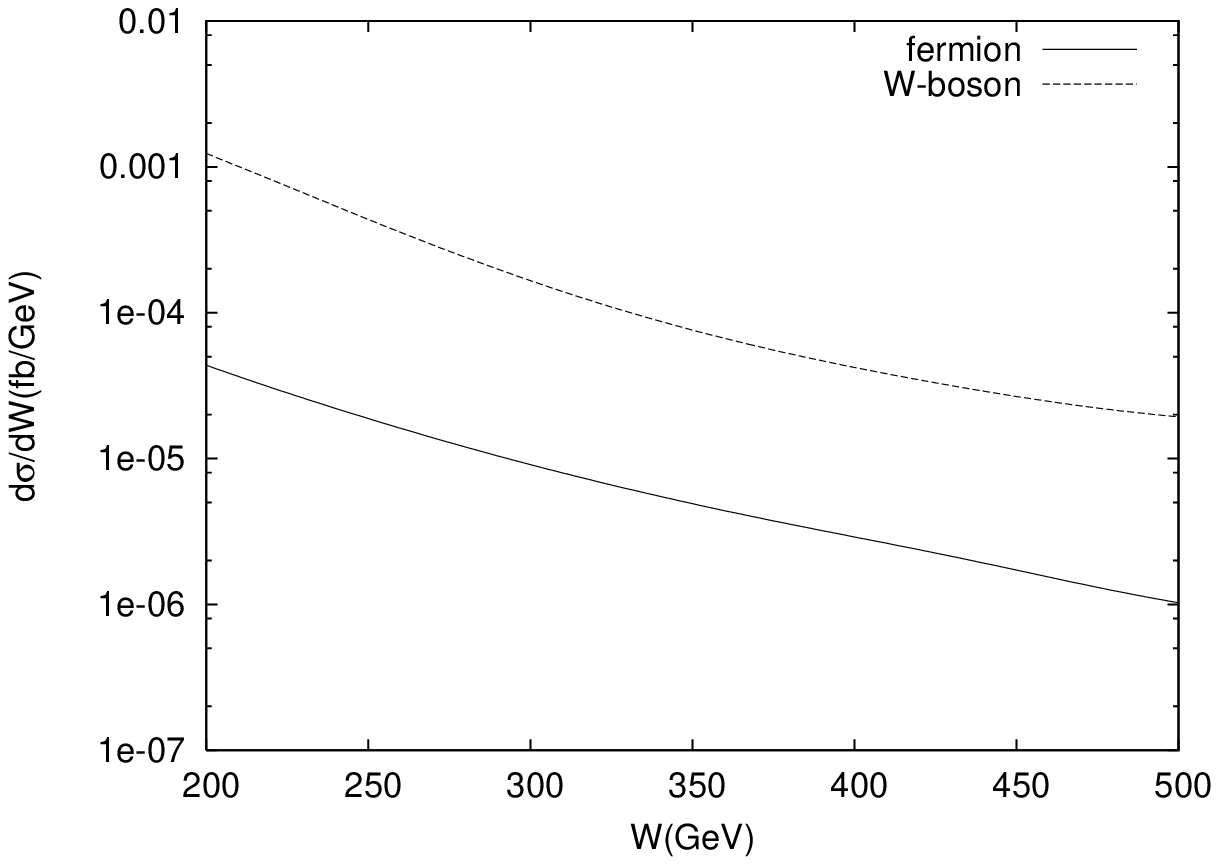}
\caption{Contributions of fermions and W boson to the
cross sections $pp \to p\gamma\gamma p$ as a function of
invariant mass of the two photon system for the energy region of
200-500 GeV and for the rapidity $|\eta|<2.5$ in the laboratory 
system. Forward dedector acceptance is taken  $0.0015<\xi<0.5$.}
\label{fig2}}

\FIGURE{\epsfig{file=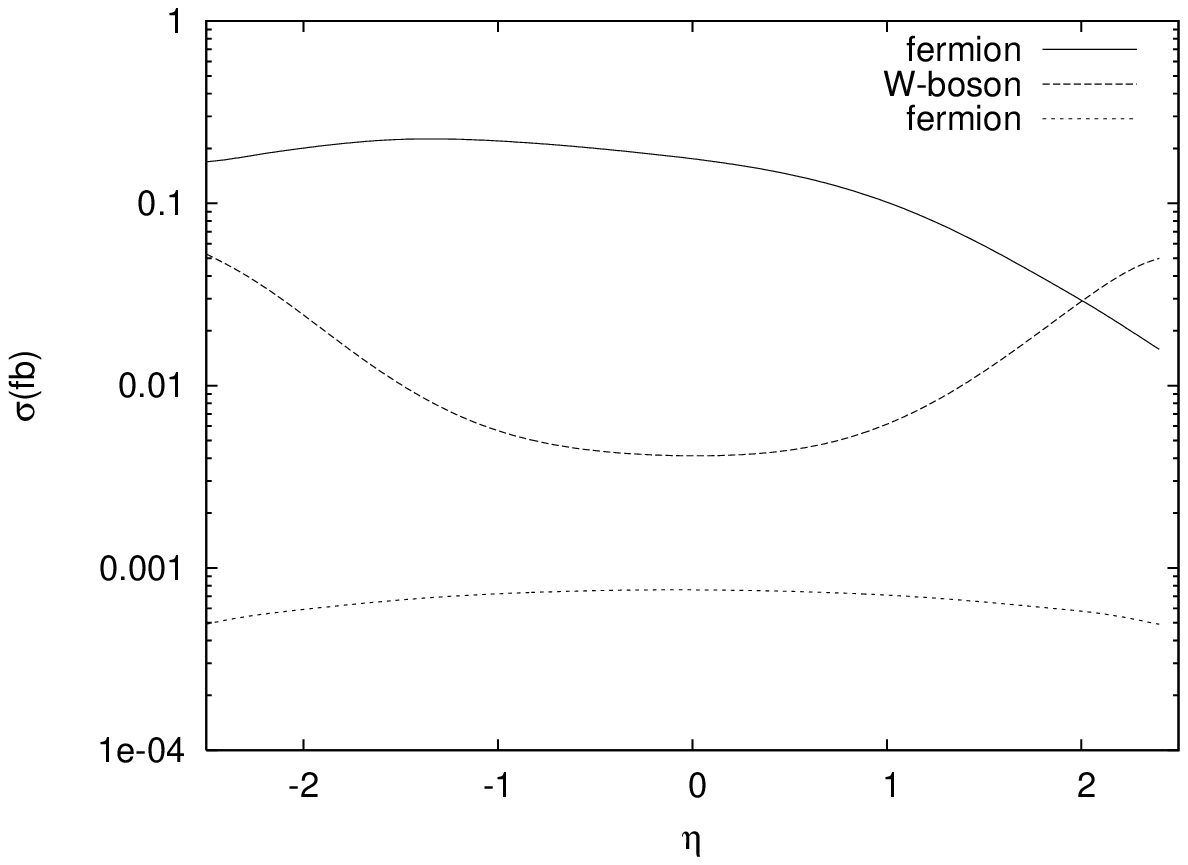}
\caption{Contributions of fermions and W boson to the
cross sections $pp \to p\gamma\gamma p$ as a function of
rapidity for the two energy regions of diphoton system    
20-80 GeV and 200-500 GeV.  
Fermion contributions are given by upper and lowest curves 
corresponding 20-80 GeV  and 200-500 GeV.
Middle curve represents W-boson contribution with 
200-500 GeV diphoton energies.
Forward dedector acceptance is taken  $0.0015<\xi<0.5$.}   
\label{fig3}}

ATLAS detector at the LHC is capable of detecting
of photons efficiently(up to 90\%) based on its wide range of 
parameters \cite{atlas}.
If sufficient luminosity of 100-200 $fb^{-1}$ is available at 
LHC, the cross sections of the  contributions of W boson and fermions 
can be observable separately in the  $\gamma\gamma \to \gamma\gamma$ 
one-loop SM process. 
All the cross sections and related distributions in this paper are 
given in the laboratory system i.e. center of mass 
system of the two incoming protons.

\section{ADD Model of Large Extra Dimensions}

The Hierarchy problem in particle
physics arises from the fact that 
 there is a large  difference 
in energy between the electroweak scale
which is a few hundreds of GeV and gravity
scale that is the Planck scale $M_{Pl}\sim 10^{19}$GeV 
in a four dimensional spacetime.
Extra space dimensions 
higher than three are already known in string theory. 
Following the string theory ideas, 
three space dimensional world is called a "wall" or 
3-brane where all Standard Model particles are  
confined to this wall. The D-dimensional spacetime,      
$D=3+\delta+1$, with $\delta$ extra space dimensions
is called "bulk" where 3-brane is embedded in it.
The methods how to handle hierarchy problem are 
model dependent.

In the model proposed by 
Arkani-Hamed, Dimopoulos and Dvali gravity propagates
in the bulk but SM fields can not go out of the 3-brane \cite{add}. 
The solutions of the 
linearized equations of motion of the metric field are 
the Kaluza-Klein tower in D dimensions. 
Upon integrations over extra 
dimensions, the final 4-dimensional fields are the 
Kaluza-Klein modes. Zero mode of the KK field is massless 
that leads to the graviton in 4-dimensions. Excited 
modes of the KK fields become  massive. 
According to this model extra dimensions are 
compactified with a compactification radius 
$r_{c}\sim $mm-fermi (or $1/r_{c}\sim $eV-MeV) 
by which  the KK mode spacing is specified. 
This mode spacing is very small with respect  to 
the typical collider energies.
This property allows for  the summation 
over large number of KK states. 
ADD model is also known as the large extra dimension model
due to large compactification radius.
Strong gravity occurs in $D=4+\delta$ dimensions because of 
the overall effect of the KK states. Thus its effective 
interactions in 4-dimensions  with the 
Standard Model particles  are expected to be  measurable at 
collider energies. The relation between  the Planck Mass 
$M_{Pl}$ and the corresponding scale $M_{D}$ in D-dimensions       
can be written by the use of the compactified volume $V_{\delta}$  

\begin{eqnarray}
M^{2}_{Pl}=V_{\delta}M^{2+\delta}_{D}
\end{eqnarray} 

When $M_{D}$ is assumed to be  in TeV 
region, large higher dimensional  volume
of $V_{\delta}$ makes  $M_{Pl}$ large  with $\delta=2-7$. 
This result  implies  
that Planck scale $M_{Pl}$ is not the fundamental scale afterall. 
The large gap between the electroweak
and Planck scale is compensated by the large
compactification scale of the extra dimensions. 
It should be pointed out that this does 
not resolve the hierarchy problem but large gap shifts to the 
compactification radius and the inverse of the electroweak scale. 

There are three Feynman diagrams containing KK propagators 
in s, t and u channels. 
Let us now calculate the Feynman amplitudes for the subprocess 
$\gamma\gamma \to \gamma\gamma$ for three channels
using the KK-$\gamma\gamma$ vertex function
$\Gamma^{\alpha\beta\rho\sigma}$ \cite{add2}  

\begin{eqnarray}
\Gamma^{\alpha\beta\rho\sigma}=-\frac{i\kappa}{2}
[(p_{1}\cdot p_{2})C^{\alpha\beta\rho\sigma}+
D^{\alpha\beta\rho\sigma}]
\end{eqnarray}
where $p_{1}, p_{2}$, $p_{3}, p_{4}$ 
are incoming and outgoing photon momenta. 
The coupling constant $\kappa$ is 
related to the Newton constant
$G^{(4+\delta)}_{N}$ in $D=4+\delta$ dimension by
$\kappa^{2}=16\pi G^{(4+\delta)}_{N}$. 
Explicit forms of the tensors $C^{\alpha\beta\rho\sigma}$ and
$D^{\alpha\beta\rho\sigma}$ are given by 

\begin{eqnarray}
C^{\alpha\beta\rho\sigma}&&=\eta^{\alpha\rho}\eta^{\beta\sigma}
+\eta^{\alpha\sigma}\eta^{\beta\rho}-
\eta^{\alpha\beta}\eta^{\rho\sigma} \\
D^{\alpha\beta\rho\sigma}&&=\eta^{\alpha\beta}
p^{\sigma}_{1}p^{\rho}_{2}-(\eta^{\alpha\sigma}
p^{\beta}_{1}p^{\rho}_{2}+\eta^{\alpha\rho}
p^{\sigma}_{1}p^{\beta}_{2}-\eta^{\rho\sigma}
p^{\alpha}_{1}p^{\beta}_{2}) \nonumber \\
&&-(\eta^{\beta\sigma}
p^{\alpha}_{1}p^{\rho}_{2}+\eta^{\beta\rho}
p^{\sigma}_{1}p^{\alpha}_{2}-\eta^{\rho\sigma}
p^{\beta}_{1}p^{\alpha}_{2})
\end{eqnarray}
where $\eta^{\rho\sigma}$ is the metric tensor of the flat 
space in four dimensions.
The square of the amplitudes are

\begin{eqnarray}
|M|^{2}&&=\frac{\kappa^{4}}{8}\{|D(\hat{s})|^{2}
(\hat{t}^{4}+\hat{u}^{4})+|D(\hat{t})|^{2}
(\hat{s}^{4}+\hat{u}^{4})+|D(\hat{u})|^{2}
(\hat{s}^{4}+\hat{t}^{4})  \nonumber \\
&&+[D^{\star}(\hat{s})
D(\hat{t})+D(\hat{s})D^{\star}(\hat{t})]
\hat{u}^{4}+[D^{\star}(\hat{s})
D(\hat{u})+D(\hat{s})D^{\star}(\hat{u})]
\hat{t}^{4} \nonumber \\
&&+[D^{\star}(\hat{t})D(\hat{u}) 
+D(\hat{t})D^{\star}(\hat{u})]
\hat{s}^{4} \}
\end{eqnarray}
Lorentz invariant Mandelstam variables $\hat{s}$, $\hat{t}$
and $\hat{u}$ are given in previous section.   
Here the factor due to initial spin average and 
statistical factor from identical final photons are absent.
From previous sections we know that
only W boson contributes to the Standard Model one-loop
amplitudes for the energies $\sqrt{\hat{s}}>200$ GeV. At these
energies the dominant terms are imaginary. This makes 
the interference terms with the SM  negligible based on the 
fact that KK contributions to the amplitudes are real.

KK propagator $D(\hat{s})$ contains
summation over Kaluza-Klein modes which can be calculated without 
specifying any specific process. Since the KK tower is an infinite 
sum, ultraviolet divergences are present in tree level process. Thus 
we need a cutoff procedure.
For the purpose of phenomenological applications basic steps 
of the approach followed by Han et al. \cite{add2}  are 
given below 

\begin{eqnarray}
\kappa^{2}D(\hat{s})&&\equiv
\kappa^{2}\sum_{n}\frac{i}{\hat{s}-m^{2}_{n}}=
\frac{8\pi \hat{s}^{\delta/2-1}}{M_{D}^{\delta+2}}[2iI(x)+\pi]
\end{eqnarray}
The last equality is obtained by approximating the infinite 
KK sum by an integral. The real part in square bracket 
in the right hand side comes from the narrow 
resonant states of single KK mode.   
$I(x)$ is related to nonresonant summation of infinite 
KK tower where $x=\frac{\Lambda_{c}}{\sqrt{\hat{s}}}$ and
$\Lambda_{c}$ is defined as the ultraviolet cutoff energy.
The definition of $\kappa^{2}=16\pi/M_{Pl}^{2}$ and Eq.(3.1)
have been used to reach the form on the right side. The 
explicit form of $I(x)$ is written for even and odd values 
of $\delta$

\begin{eqnarray}
I(x)&&=-\sum_{k=1}^{\delta/2-1}\frac{1}{2k}x^{2k}
-\frac{1}{2}\log{(x^2-1)} \;\;\;\;\; \;\; (\delta=\mbox{even}) \\
&&=-\sum_{k=1}^{(\delta-1)/2}\frac{1}{2k-1}x^{2k-1}
+\frac{1}{2}\log{(\frac{x+1}{x-1})}\;\;\;\;\; \;\; 
(\delta=\mbox{odd})  
\end{eqnarray}

The relation between the unknown cutoff energy 
$\Lambda_{c}$ and the fundamental scale
$M_{D}$ is not clear unless the full theory is known.
The connection $\Lambda_{c} < M_{D}$ can be given on the ground
of  the string theory. In several calculations in the 
literature  the equality
$M_{D}\simeq \Lambda_{c}$ demonstrates the lower limit
of the fundamental scale $M_{D}$.
Within this work $M_{D}\simeq \Lambda_{c}$ is set for KK graviton
propagator of ADD model.
$\kappa^{2}D(\hat{s})$ is dominated by nonresonant ultraviolet 
contribution for $x>>1$. Then approximate result can be obtained 
for comparison with  other notations in the literature \cite{add2}. 

\begin{eqnarray}
\kappa^{2}D(\hat{s})=-\frac{16i\pi}{(\delta-2)\Lambda_{c}^{4}}
\;\;\;\;\; \;\; (\delta > 2)
\end{eqnarray}
This corresponds to the result of Giudice et al. \cite{add2}  
$\Lambda_{c}=\Lambda_{T}$ for $\delta=4$. 
The same form can be used for t and u channels. 

Collider signals of virtual graviton exchange manifest itself 
as the deviations from the SM in the cross section. Furthermore,  
the angular distributions of the final particles
are dedicated to spin-2 nature of the graviton. 
First we show  the rapidity distribution of
the  final photons in Fig.\ref{fig4} for the  contribution 
of the KK graviton with $M_{D}=1500$ GeV 
 to the total cross section of the main process 
$pp \to p \gamma\gamma p$ in the photon induced interactions 
at LHC with forward detectors. We consider two acceptance 
regions of the  forward detectors
$0.0015<\xi<0.5$ and $0.1<\xi<0.5$. Each curve correspons to 
a rapidity cut $|\eta|<2$  with  additional 
cuts on variables  $\sqrt{\hat{s}}>200$ GeV, 
$\sqrt{|\hat{t}|}>200$ GeV and $\sqrt{|\hat{u}|}>200$ GeV. 
Requirement  both $\sqrt{|\hat{t}|}>200$ and 
$\sqrt{|\hat{u}|}>200$ GeV induces a cut  $|\eta|<2$.  
We did not include SM one-loop contribution 
 on the ground of its smallness that can be seen 
in Fig.\ref{fig3}. All cuts are also given in the 
laboratory frame of the protons as stated earlier

\FIGURE{\epsfig{file=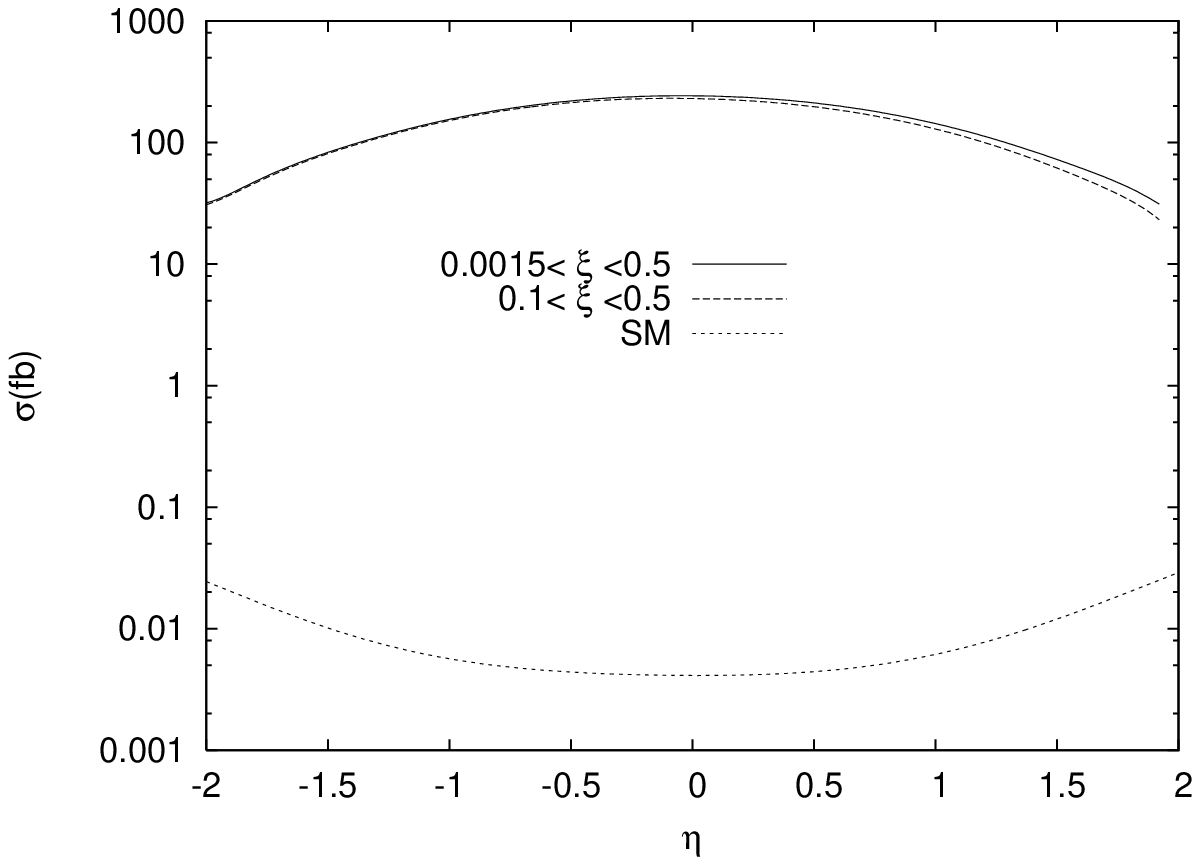}
\caption{Rapidity distributions of the KK terms in ADD model 
for the rapidity cut $|\eta|<2$, $M_{D}=1500$ GeV  
with two acceptance regions $0.0015<\xi<0.5$ and
$0.1<\xi<0.5$. Cuts and results are given in the 
laboratory frame of the protons.}
\label{fig4}}

$p_{T}$ distribution of the final photons for the KK terms of the 
cross section is plotted in Fig.\ref{fig5} with $|\eta|<2$ and 
two acceptance regions $0.0015<\xi<0.5$ and $0.1<\xi<0.5$.
In Fig.\ref{fig4} and Fig.\ref{fig5} both curves for 
different acceptance regions  seem  almost
identical in  shape and in the magnitude.
This leads to the fact that the cuts on   
$\sqrt{|\hat{t}|}$ and $\sqrt{|\hat{u}|}$ are satisfied by 
high $\sqrt{\hat{s}}$ values.
It is clear that the contribution of the energy from  
the region $0.0015<\xi<0.1$ to the cross section
of the  KK terms  is negligible for the cuts given above. 
Thus, it is more convenient to use the region $0.1<\xi<0.5$
for the rest of our calculations.
In this energy region cross section  from SM one-loop 
contributions is the order of $10^{-4}$fb  where we ignore
it  from here on.

\FIGURE{\epsfig{file=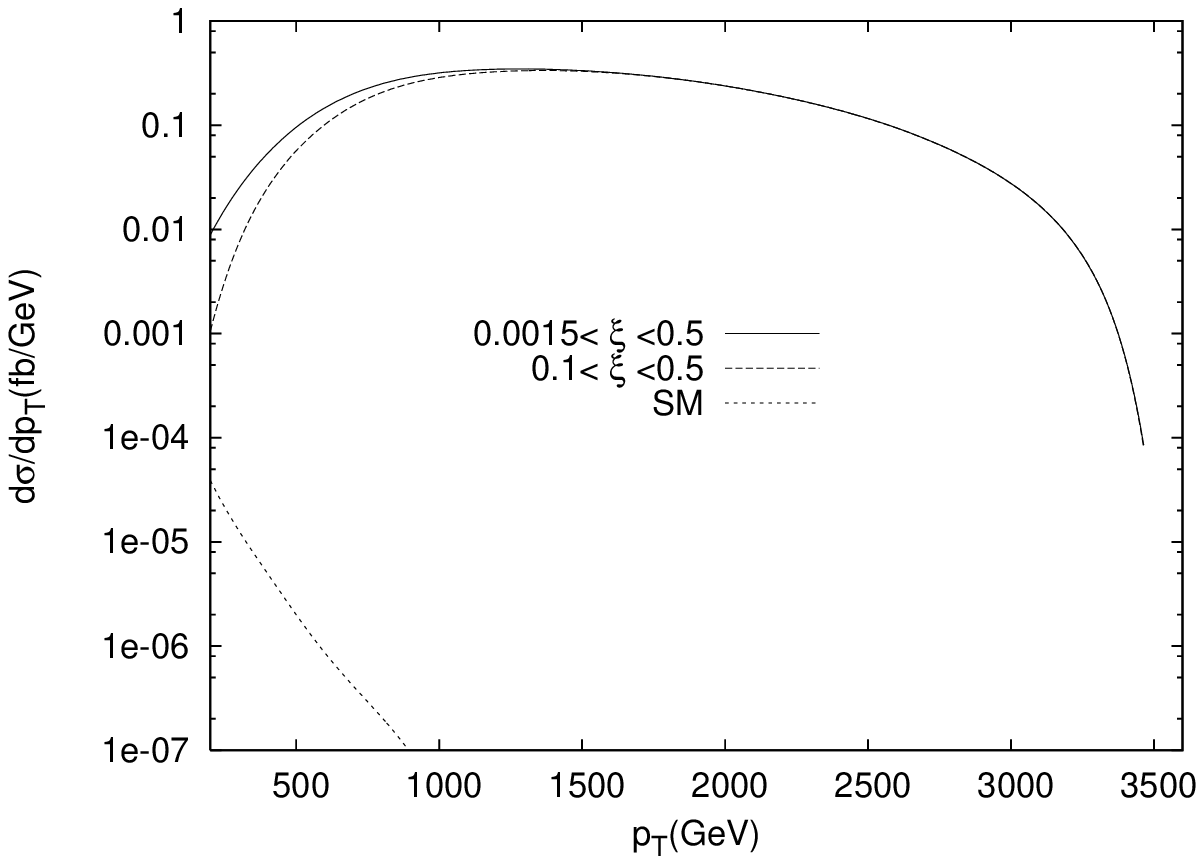}
\caption{$p_{T}$ distribution of the final photons for the KK 
graviton exchange terms in ADD model with the same values of the 
parameters  $|\eta|$, $\xi$  and $M_{D}$ as in 
the previous figure.} 
\label{fig5}}

Next, we calculate 95\% C.L.  bounds
on the $M_{D}$ as a function of the integrated LHC luminosity
for  the acceptance region $0.1<\xi<0.5$ with   
the above restrictions on the rapidity and the Mandelstam variables.
Poisson distributed events are considered for statistical 
analysis. The estimations are shown in Fig.\ref{fig6} where 
the acceptance region $0.1<\xi<0.5$ is 
good enough to feel the KK graviton exchange in the process
$pp \to p\gamma\gamma p$ via two photon fusion.

\FIGURE{\epsfig{file=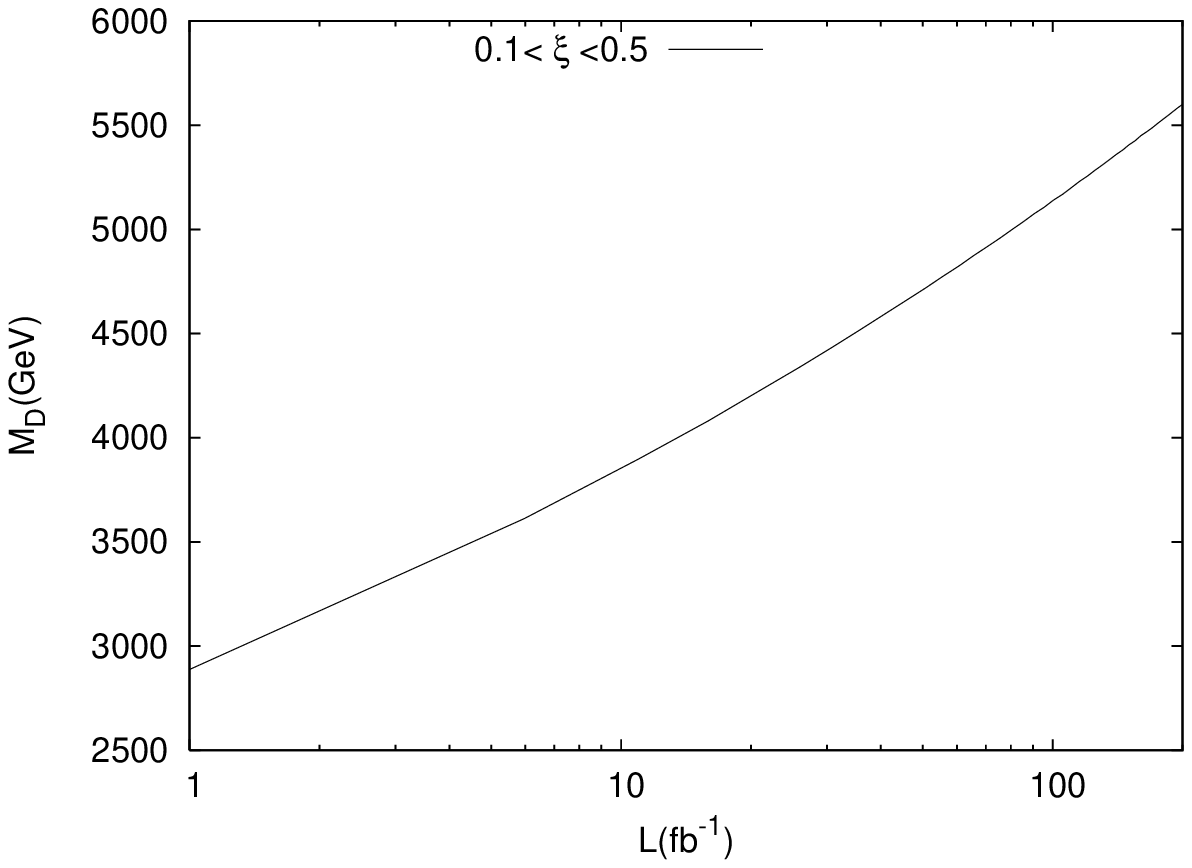}
\caption{95\% C.L. search reach for $M_{D}$ as a function of 
integrated LHC luminosity for the acceptance region
$0.1<\xi<0.5$  and rapidity cut $|\eta|<2$ in ADD model.}
\label{fig6}}

\section{RS Model of Warped Extra Dimensions}

Randall and Sundrum proposed a model to handle 
the large  hierarchy between the electroweak
and the gravity scales. In this model curvature in higher dimensions
play the important role to  remove this large hierarchy.
The metric in 5D is  a solution
to Einstein's equations respecting the 4D Poincare invariance \cite{rs} 

\begin{eqnarray}
ds^{2}=e^{-2ky}\eta_{\mu\nu}dx^{\mu}dx^{\nu}-dy^{2},
\end{eqnarray}     
where y is the $5^{th}$ dimension which is parametrized as
$y=r_{c}|\Phi|$ with $r_{c}$ being the compactification radius
of the extra dimension. Angular coordinate $\Phi$
has the limits of $0\leq|\Phi|\leq\pi$.
First term in this metric is the Minkowski spacetime 
metric in 4D  multiplied by an exponential factor which 
is called warp factor containing fifth 
dimension and the degree of the curvature k. 
This model includes two 3-brane with 
opposite and equal tensions  at the boundaries 
of  a 5D  Anti-de-Sitter space. 
Each boundary has 4D Minkowski 
metric and y is orthogonal to each 3-brane. 
The distance between the two walls is $y=\pi r_{c}$.
The wall at $y=0$ on which  Gravity is localized
is called Planck brane. The other wall at $y=\pi r_{c}$ 
is referred to TeV brane where SM fields live on. 
Gravity propagates in $5^{th}$ dimension.

Starting with 5D action, the 4D effective
lagrangian containing  the interaction of the KK gravitons
with the matter fields can be obtained by

\begin{eqnarray}
L=-\frac{1}{\bar{M}_{Pl}}T^{\alpha\beta}(x)h^{(0)}_{\alpha\beta}(x)
-\frac{1}{\Lambda_{\pi}}T^{\alpha\beta}(x)
\sum_{n=1}^{\infty}h^{(n)}_{\alpha\beta}(x)
\end{eqnarray}
where $T^{\alpha\beta}(x)$ is the energy momentum tensor of the 
matter field in the Minkowski space and 
$\bar{M}_{Pl}=M_{Pl}/\sqrt{8\pi}$ is the reduced Planck scale.
$h^{(n)}_{\alpha\beta}(x)$ describes the KK modes of 
the graviton on the 3-brane. The masless 
zero KK mode decouples from the sum and its coupling 
strength is $1/\bar{M}_{Pl}$. The massive  KK states has the 
coupling of $(1/\Lambda_{\pi})\sim 1/\mbox{TeV}$ with 
$\Lambda_{\pi}=e^{-kr_{c}\pi}\bar{M}_{Pl}$. 

The  relation between 5D fundamental Planck scale M and 
usual 4D reduced Planck scale is  

\begin{eqnarray}
\bar{M}^{2}_{Pl}=\frac{M^{3}}{k}(1-e^{-2kr_{c}\pi}).
\end{eqnarray}    
Taking  $k\sim M_{Pl}$  we reach  the 
relation $M\sim M_{Pl}$. Therefore, there is no 
additional hierarchy created by the model. If 
$kr_{c}$ has the value  $\sim 10-12$  all the physical 
processes take place in TeV scale on TeV-brane. This  
demonstrates the fact  that  the hierarchy is 
generated by the warp factor.  

The mass spectrum created by  KK modes of the graviton
in the RS model  is given by \cite{rs} 

\begin{eqnarray}
m_{n}=x_{n}ke^{-kr_{c}\pi}=x_{n}\beta\Lambda_{\pi}, 
 \;\; \mbox{with}\;\; \beta=\frac{k}{\bar{M}_{Pl}}
\end{eqnarray}
where $x_{n}$ are the roots of the Bessel function of order 1
$J_{1}(x_{n})=0$. The first values are $x_{1}=3.83$, $x_{2}=7.02$
and $x_{3}=10.17$. The scale of the  masses is   
$m_{n}\sim \Lambda_{\pi}\sim \mbox{TeV}$
on the ground of   $\beta\sim 1$. It is clear   that
the each excitation should be sizable separately
at colliders. The appropriate parameters to describe 
the RS scenario for phenomenological applications 
are $\Lambda_{\pi}$ and $\beta$.
The only graviton propagator  differs  from the ADD case 
in the squared amplitude for $\gamma\gamma \to \gamma\gamma$

\begin{eqnarray}
\kappa^{2}D(\hat{s})=\frac{2}{\Lambda^{2}_{\pi}}
\sum_{n=1}^{\infty}\frac{1}{\hat{s}-m^{2}_{n}+i\Gamma_{n}m_{n}},
\;\;\;\;\;  \Gamma_{n}=\rho m_{n}(\frac{m_{n}}{\Lambda_{\pi}})^{2}
\end{eqnarray}  
with $\rho=1$ is used in the width $\Gamma_{n}$ of the individual 
KK graviton. 
Limits on the parameter $\beta$ can be found for the first graviton 
mode  with mass $m_{1}$. The estimation for 95\% C.L. parameter 
exclusion region is shown in Fig.\ref{fig7} for the integrated LHC 
luminosities; $50 fb^{-1}$, $100 fb^{-1}$ and $200 fb^{-1}$,
with the acceptance region $0.1 <\xi < 0.5$. 

\FIGURE{\epsfig{file=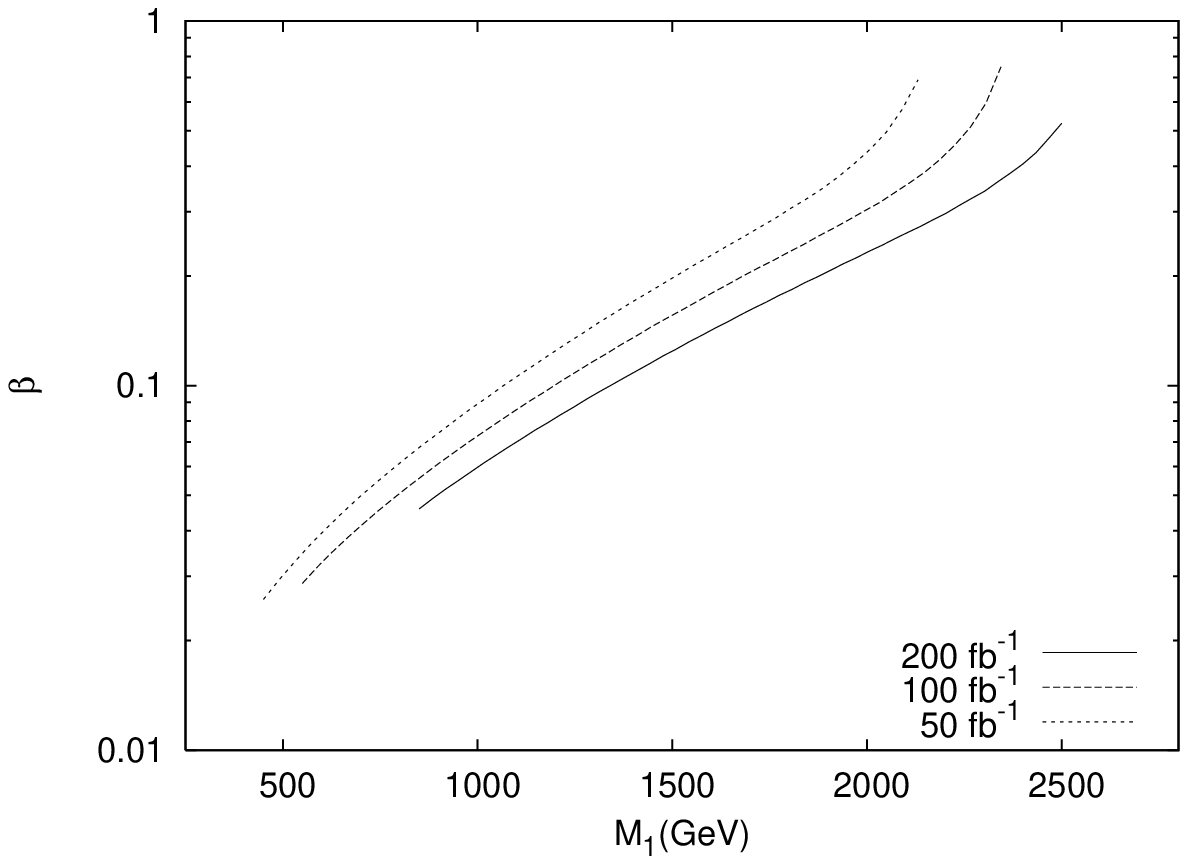}
\caption{95\% C.L. exclusion region for the parameters 
$\beta$ and $m_{1}$ in RS model  at three different integrated LHC 
luminosities; $50 fb^{-1}$, $100 fb^{-1}$ and $200 fb^{-1}$.
Other cuts are the same as in the previous figure.
Excluded regions are defined by the area over the curves.}
\label{fig7}}

In the approximation $m^{2}_{n}>>\hat{s}$, $m^{2}_{n}>>\hat{t}$,
$m^{2}_{n}>>\hat{u}$  KK graviton propagators for  three channels
take the form

\begin{eqnarray}
\kappa^{2}D(\hat{s})=\frac{2}{\beta^{2}\Lambda^{4}_{\pi}}
\sum_{n=1}^{\infty}\frac{-1}{x^{2}_{n}}.
\end{eqnarray}

Fig.\ref{fig8} shows 95\% C.L. search reaches 
in the $\Lambda_{\pi} - \beta$ plane for an acceptance region of 
$0.1 <\xi<0.5$ and LHC luminosities given before.

\FIGURE{\epsfig{file=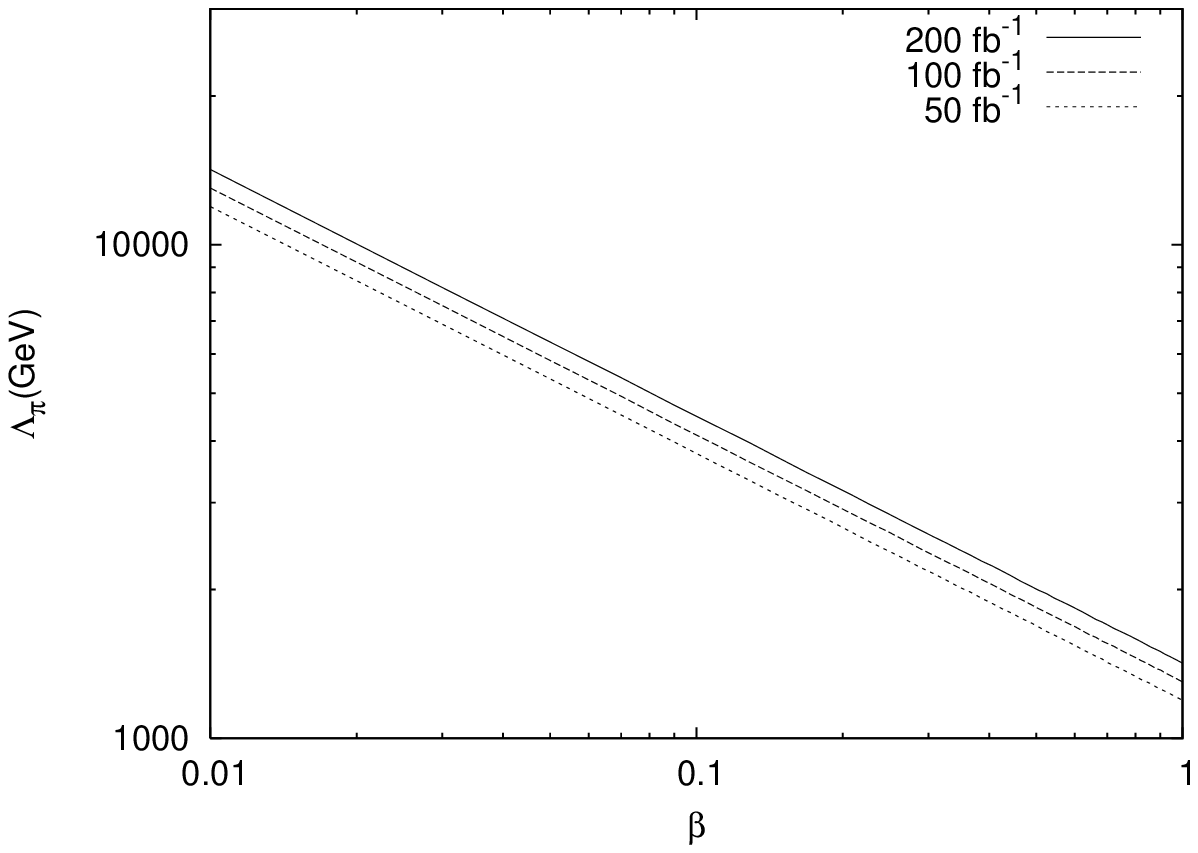}
\caption{95\% C.L. constraints  on the parameter plane
$\Lambda_{\pi}$ and $\beta$ in RS model 
 at three different integrated LHC
luminosities; $50 fb^{-1}$, $100 fb^{-1}$ and $200 fb^{-1}$
and the acceptance region $0.1 <\xi<0.5$.
Excluded regions are defined 
by the area below the curves.}
\label{fig8}}

\section{Conclusion}

Photon-photon collision at LHC with  diphoton invariant mass 
$ W> 1$ TeV allows to study physics at TeV scale beyond 
the SM  with a sufficient luminosity. 
There is no existing collider with this property except the 
LHC itself. It is worth mentioning that 
two photon final state is one of the cleanest channel to search 
for any deviation from the SM physics. For this reason,
we have investigated how the photon-induced two photon final 
state can extend the bounds on the model parameters of 
extra dimensions in the framework of the ADD and RS models.
Taking an acceptance region of $0.1 < \xi < 0.5$ 
we have obtained constraints on the fundamental 
scale $M_{D}$ in the ADD model for a LHC
luminosity interval 1-200 $fb^{-1}$.  
Exclusion regions of the parameter 
pairs of RS model $\beta -m_{1}$ and 
$\Lambda_{\pi}-\beta$  have been provided for the LHC 
luminosities $50 fb^{-1}$, $100 fb^{-1}$ and $200 fb^{-1}$.
Possible background coming from the SM one-loop diagrams can 
be neglected  in the energy and rapidity range defined above 
sections.
Excluded area of the model parameters that we have found from the 
process $pp \to p\gamma\gamma p$ extends to wider regions than the 
case of the   colliders LEP and Tevatron  \cite{rizzo,han}. 
Certainly, the challenging processes are the quark antiquark annihilation 
into two photons $q\bar{q} \to \gamma\gamma$ as well as  gluon fusion
$gg \to \gamma\gamma$ with KK graviton exchange at the LHC itself
\cite{han}. Although $\gamma\gamma \to \gamma\gamma$ process 
has cleaner enviroment, $gg, q\bar{q} \to \gamma\gamma$ process 
needs a sophisticated   background analysis. 
The excluded regions  of this paper are wider than the case of 
the similar study using the subprocess 
$\gamma\gamma \to \ell^{+}\ell^{-}$ at LHC \cite{atag}.

\end{document}